\documentclass[12pt]{article}

\catcode`\@=11 \@addtoreset{equation}{section}\catcode`\@=12

\begin{document}
\baselineskip8mm

 \title{\vspace{-2cm}
 Dually-charged mesoatom on the space of constant negative curvature} 
 \author{V. D. Ivashchuk$^{\dag}$ and V. N. Melnikov  }
 \date{}
 \maketitle
 {\em Center for Gravitation and Fundamental Metrology, VNIIMS, 3/1
 M. Ulyanovoy str., Moscow, 119313, Russia}\\

The discrete spectrum solutions corresponding to dually-charged
mesoatom on the space of constant negative curvature are obtained.
The discrete spectrum of energies is finite and vanishes, when
the magnetic charge of the nucleus exceeds the critical value. \\


$^{\dag}$e-mail: vivashchuk@sci.pfu.edu.ru 

\pagebreak 

\section{Introduction}

The behaviour of atom-like systems in curved backgrounds were studied
by many authors (see, for example, 
\cite{A}-\cite{IM} and references cited there).
A lot of papers were devoted to calculations of the curvature-induced 
energy-level shifts within the framework of the perturbation theory.

In this paper we consider the "motion" of massive charged  scalar
particle (meson) in the field of static dually-charged nucleus on
the space of constant negative curvature. We find the discrete 
spectrum solutions of the Klein-Gordon equation (see formulas (3.29)
and (3.31)). The discrete spectrum of the mesoatom is finite. The
largest principle number $N_{0}$ (see (3.26)) depends on the radius
of curvature $a$ and the magnetic charge $g_{m}$. For sufficiently 
small values of $a$ or large values of $g_{m}$ the discrete spectrum
is empty.

It should be noted that the expression for the energy levels (formula
(3.24) of this paper) was obtained earlier in \cite{S}. But the 
expressions for $N_{0}$ and the wave  functions in \cite{S} are wrong 
\cite{IM}.

\section{The model}

We consider the space-time $R \times {L^{3}}(a)$ with the metric
\begin{equation}
g=c^{2}dt \otimes dt -\gamma= {g_{\mu \nu}}(x) dx^{\mu} \otimes dx^{\nu},
\end{equation}
where
\begin{equation}
{L^{3}}(a) \equiv \{z|z = (z^{0}, \vec{z}) \in R^{4}, \, z^{0} > 0, \,
  (z^{0})^{2} - (\vec{z})^{2} = a^{2} \} 
\end{equation}
is 3-dimensional space of constant negative curvature ($a$ is radius of 
curvature) with the canonical metric
\begin{equation}
\gamma = {\gamma_{ij}}(\vec{x})dx^{i} \otimes dx^{j}=
 a^{2}[d \chi \otimes d \chi + \sinh^{2} \chi (d \theta \otimes d \theta
+ \sin^{2} \theta d \varphi \otimes d \varphi)],
\end{equation} $0 < \chi < + \infty$ \, ($a \sinh \chi = |\vec{z}|)$.

We consider a static dually-charged nucleus with the electric charge 
$(-Ze)$ and a magnetic charge $g_{m}$, placed in the coordinate origin 
$\chi = 0$. Let $U \subset {L^{3}}(a)$ be a domain with the trivial 
cohomology group ${H^{2}}(U,R)=0$ and $\{\chi = 0\} \not \in U$.
The electromagnetic 4-potential $A_{\mu}$ on $R \times U$, 
corresponding to the nucleus has the following form
\begin{equation}
A = A_{\mu}dx^{\mu} = (- \frac{Ze}{a})(\coth \chi -1)dt + {\cal A},
\end{equation}
where
\begin{equation}
{\cal F} =d{\cal A} = g_{m} \sin \theta d\theta \wedge d\varphi
\end{equation}
is the strength of the electromagnetic field, corresponding to the magnetic
charge $g_{m}$. The relation (2.5) is correct, since due to $H^{2}(U,R) =0$
any closed 2-form on $U$ is exact, i.e. $d {\cal F} = 0$ entails the 
existence of ${\cal A}$ such that $d {\cal A} = {\cal F}$. For $U = U_{\pm}$
, where
\begin{equation}
U_{\pm} = {L^{3}}(a) \setminus \{\theta = \frac{\pi}{2} \pm  
\frac{\pi}{2}\}, \nonumber \end{equation} the 1-form on  $U = U_{\pm}$ 
\begin{equation}
{\cal A} = {\cal A}^{\pm} = g_{m}(\pm 1 - \cos \theta) d \varphi
\end{equation}
satisfy the relation (2.5).

A massive charged scalar particle (meson), moving in the field of the 
static dually--charged nucleus, has the following action
\begin{equation}
S[\varphi] = \frac{1}{c} \int_{M_{\ast}} d^{4}x (-\det g_{\mu \nu})
^{1/2}\{\hbar^{2} g^{\mu \nu} (\overline{D_{\mu}(A^{\ast})\varphi_{\ast}})
(D_{\nu}(A^{\ast})\varphi_{\ast}) - m_{0}^{2}c^{2}\bar{\varphi_{\ast}} 
\varphi_{\ast}\}, \end{equation}
where $D_{\mu} = {D_{\mu}}(A^{\ast}) \equiv \nabla_{\mu} + (ie/
\hbar c) A^{\ast}_{\mu}$, $\nabla_{\mu}$ is covariant derivative,
corresponding to the metric (2.1); the  symbol $\ast = \pm$ and
$A= A^{\pm}$ is a result of substitution of ${\cal A}^{\pm}$
from (2.7) to (2.4); $m_{0}$  is mass of the scalar particle and $e$    
is its charge (opposite in sign to the nucleus charge). The pair of 
functions
\begin{equation}
 \varphi_{\pm} : M_{\pm} = R \times U_{\pm} \longrightarrow C
\end{equation}
satisfy the overlapping condition
\begin{equation}
{\varphi_{+}}(t,\vec{x}) = {\Omega}(\vec{x}){\varphi_{-}}(t,\vec{x}),
\end{equation}
$\vec{x} \in U_{+} \cup U_{-}$, where
\begin{equation}
\Omega : U_{+} \cup U_{-} \longrightarrow  U(1)
\end{equation}
is a smooth overlapping function. The scalar particle (meson) wave function
is a smooth section of a vector $C$-bundle  with the base  $R \times
({L^{3}}(a) \setminus \{\chi =0\})$. This section is defined by the pair 
of functions (2.9), satisfying the condition (2.10). 
(The function $\varphi_{\pm}$ is the representation of the function 
$\varphi | M_{\pm}$ in the local trivialization over $M_{\pm}$).

The action (2.8) is correctly defined, i.e. the right hand side of (2.8)
does not depend on the choice of the symbol $\ast = \pm$ (or equivalently 
on the choice of local trivialization) if the function $\Omega$ 
(2.11) satisfies the following relation on $U_{+} \cup U_{-}$
\begin{equation}
{\cal A}^{+} = {\cal A}^{-} + i \frac{\hbar c}{e} \Omega^{-1} d  \Omega
\end{equation}
(${\cal A}^{\pm}$ are defined in (2.7)). It follows from the relations 
(2.7) and (2.12), that such function does exist if and only if the Dirac 
quantization condition is satisfied \cite{D}
\begin{equation}
q \equiv e g_{m}/ \hbar c = 0, \pm \frac{1}{2}, \pm \frac{3}{2}, \ldots . 
\end{equation}
In this case
\begin{equation} \Omega = exp[-2iq(\varphi - \varphi_{0})],
\end{equation}
where  $\varphi_{0} =const$.

Varying the action (2.8), we obtain the following equation of motion
\begin{equation}
[\hbar^{2} g^{\mu \nu} (\overline{{D_{\mu}}(A^{\ast})})({D_{\nu}}(A^{\ast}))
+ m_{0}^{2}c^{2}] \varphi_{\ast} = 0.
\end{equation}

The Lagrangian, corresponding to the action (2.8), has the following 
form \begin{eqnarray} L(\varphi, v) =&&\int_{U_{\ast}} d^{3}\vec{x}(det 
\gamma_{ij})^{1/2} \{\frac {\hbar^{2}}{c^{2}}|v_{\ast} +  
\frac {ie}{\hbar} V \varphi_{\ast}|^{2}- \nonumber \\ 
&&\hbar^{2} \gamma^{ij } (\overline{{D_{i}}({\cal A}^
{\ast})\varphi_{\ast}}) ({D_{j}}({\cal A}^{\ast})\varphi_{\ast}) 
- m_{0}^{2}c^{2}\bar{\varphi_{\ast}} 
\varphi_{\ast}\},
\end{eqnarray}
where $V \equiv (-Ze)(\coth \chi -1)/a$, ${v_{+}}(\vec{x}) = 
\Omega {v_{-}}(\vec{x})$. The Lagrangian (2.16) is a continuous mapping
\begin{equation}
L : H \times H \longrightarrow R,
\end{equation}
where $H \times H \cong TH$ and $TH$ is tangent vector bundle over 
the Hilbert space $H$. This Hilbert space is the configuration space 
of the Lagrange system. It consists of smooth sections of the 
monopole vector $C$- bundle over ${L^{3}}(a) \setminus \{\chi =0 \}$ 
satisfying the restriction
\begin{equation}
\int_{U_{\ast}} d^{3}\vec{x}(det \gamma_{ij})^{1/2}\{ 
\bar{\varphi_{\ast}} \varphi_{\ast} (1 + V^{2}) +
\gamma^{ij } \overline{({D_{i}}({\cal A}^{\ast})\varphi_{\ast})}
({D_{j}}({\cal A}^{\ast})\varphi_{\ast})\} < + \infty .
\end{equation}
The scalar product in $H$ is the following
\begin{equation}
(\psi,\varphi) \equiv \int_{U_{\ast}} d^{3}\vec{x}(det \gamma_{ij})^{1/2} 
\{ \bar{\psi_{\ast}} \varphi_{\ast} (1 + V^{2}) +
\gamma^{ij } \overline{({D_{i}}({\cal A}^{\ast}) \psi_{\ast})}
({D_{j}}({\cal A}^{\ast})\varphi_{\ast}) \}
\end{equation}
$\ast = \pm$. Strictly speaking, $H$ is the completion of the 
pre-Hilbert space (with scalar product (2.19)) of smooth sections with 
compact support in $U_{+} \cup U_{-}$. ($H$ is the modified Sobolev space.) 
The field equation (2.15) is equivalent to the Euler-Lagrange equations 
for the Lagrange system $(L,H)$.

\section{The discrete spectrum solutions}

We seek solutions of the equation of motion (2.15) in the following form
\begin{equation}
{\varphi}(t,\vec{x}) = \exp (-iEt/ \hbar ) {F}(\vec{x}),
\end{equation}
where $E \in C$ and $F \in H$. The substitution of (3.1) into (2.15) 
leads to the following relation
\begin{equation}
\{[\varepsilon + Z \alpha (coth \chi -1)]^{2} + \frac{1}{\sinh^{2}\chi }
\frac {\partial}{\partial \chi}(\sinh^{2} \chi 
\frac {\partial}{\partial \chi}) + \frac{1}{\sinh^{2}\chi } 
\triangle^{\ast}_{q} - \mu^{2}\} F_{\ast} = 0,
\end{equation}
where  
\begin{equation}
\varepsilon \equiv Ea/ \hbar c, \qquad \mu \equiv m_{0}ac/\hbar, \qquad 
\alpha \equiv e^{2}/ \hbar c, 
\end{equation}
and                                
\begin{equation}    \triangle^{\ast}_{q} = \beta^{ij}
{D_{i}}({\cal A}^{\ast}){D_{j}}({\cal A}^{\ast})  \nonumber
\end{equation}
is the "monopole Laplace operator" \cite{T}
on 2-dimensional sphere $S^{2}$ (
$\beta$ is the canonical metric on $S^{2}$), written in the local 
trivialization over $S^{2}_{\ast}$, $\ast = \pm$, where $S^{2}_{\pm}
= S^{2} \setminus \{\theta = \frac {\pi}{2} \pm \frac {\pi}{2}\}$.                              
The operator $\triangle_{q}$ acts on the sections of the monopole 
vector  $C$-bundle over $S^{2}$. For $q=0$ it coincides with the Laplace 
operator on $S^{2}$. The spectrum of $\triangle_{q}$ is well-known 
\cite{T,WY}, it is discrete
\begin{equation}
\triangle_{q} Y_{qlm} = [-l(l+1) + q^{2}]Y_{qlm},
\end{equation}
where
\begin{equation}
l = |q|, |q| + 1, \ldots ; \qquad m = -l, -l+1, \ldots , l ;
\end{equation}
and $Y_{qlm}$ are monopole spherical harmonics [13]. For the sake of 
completeness  the explicit expression for $Y_{qlm}$ is presented in the
Appendix. The relation (3.5) follows from the representation for
$\triangle_{q}$ \cite{WY}
\begin{equation}
-\hbar^{2} \triangle_{q}  = (\vec{L}_{q})^{2} - \hbar^{2} q^{2}.
\end{equation}
In (3.7) $\vec{L}_{q}$ is the modified (monopole) momentum operator \cite{WY}
\begin{equation}
(L_{q}^{j})^{\ast} = \varepsilon _{jkl} z^{k}(-i \hbar 
\frac {\partial}{\partial z^{l}} +\frac{e}{c} {\cal A}_{l}^{\ast})
- \hbar q \frac{z^{j}}{|z|},
\end{equation}
$j = 1, 2, 3$; where ${\cal A}_{i}^{\pm}$ are the components of the 
1-form (2.7) in $z$-coordinates (see (2.2))
\begin{equation}
{\cal A}^{\pm}  =  {\cal A}_{i}^{\pm} dz^{i}= \frac
{g_{m} \varepsilon _{ij3} z^{i} dz^{j}}{|z|(z^{3} \pm |z|)}.   
\end{equation}
The components of the operator (3.8) satisfy the commutation relations
\begin{equation}
[L_{q}^{k}, L_{q}^{l}] = i \hbar \varepsilon_{klj} L_{q}^{j}. \nonumber
\end{equation}
The monopole harmonics $Y_{qlm}$  form a complete orthonormal set 
(on $S^{2}$) of the eigenfunctions of the operators $(\vec{L}_{q})^{2}$ 
and $L_{q}^{3}$:
\begin{eqnarray}
&[(\vec{L}_{q})^{2} - \hbar^{2} l(l+1)] Y_{qlm} = 0, \\
&[L_{q}^{3} - \hbar m ]Y_{qlm} = 0,
\end{eqnarray}
where $l$ and $m$ satisfy (3.6). The equality (3.5) follows from the 
relations (3.7) and (3.10).

Let $F$ be an eigenfunction of the operators $(\vec{L}_{q})^{2}$  
and $L_{q}^{3}$. Then                                          
\begin{equation}
{F_{\ast}}(\chi, \theta, \varphi) = {Q}(\chi){(Y_{qlm})_{\ast}}(
\theta, \varphi).
\end{equation}
Substituting (3.13) into (3.2) and taking into account (3.5), we get
\begin{eqnarray}
\{[\varepsilon + Z \alpha (coth \chi -1)]^{2} + &&\frac{1}{\sinh^{2}\chi }
\frac{\partial}{\partial \chi}(\sinh^{2} \chi 
\frac{\partial}{\partial \chi}) \nonumber \\
&&- \frac{1}{\sinh^{2}\chi} 
[l(l+1) - q^{2}] - \mu^{2} \} Q = 0,
\end{eqnarray}
The inclusion $F \in H$ is equivalent to the convergence of the integral
\begin{equation}
\int_{0}^{\infty} d \chi \, \sinh^{2}\chi \{|Q|^{2}(1+
\frac{1}{\sinh^{2}\chi }) + |\partial_{\chi} Q|^{2}\} < +\infty
\end{equation}
(this condition follows from (2.18) and (3.13)).

We introduce a new variable $x$
\begin{equation}
x = 2/(\coth \chi +1)
\end{equation}
($0 < x < 1$ for $\chi > 0$). Then  eq. (3.14), written 
in $x$-variable,
\begin{eqnarray}
\frac{d^{2}Q}{dx^{2}} + \frac{2}{x} \frac{dQ}{dx}  + &&\frac{1}
{4x^{2}(1-x)^{2}} \{[\varepsilon x +2Z \alpha (1-x)]^{2} 
\nonumber \\
&&- \mu^{2}x^{2} - 4[l(l+1)-q^{2}](1-x)]Q = 0
\end{eqnarray}
has a generalyzed-hypergeometric form \cite{NU}. The standard procedure (see,
for example [14]) give the substitution
\begin{equation}
Q = x^{- \frac{1}{2} + \kappa}(1-x)^{\frac{1}{2} + \frac{\lambda}{2}}v,
\end{equation}
leading to the hypergeometric equation for the function $v = {v}(x)$
\begin{eqnarray}
x(1 - x)\frac{d^{2}v}{dx^{2}} + &&[1 + 2\kappa  - (2+ 2 \kappa +
\lambda)x] \frac{dv}{dx} +  \nonumber \\
&&[Z \alpha \varepsilon -
(\kappa + \frac{1}{2})^{2}  - (Z \alpha)^{2}
- \lambda (\kappa + \frac{1}{2})]v = 0,
\end{eqnarray}
where
\begin{equation}
\lambda = \sqrt{\mu^{2} + 1 - \varepsilon^{2}}, \qquad
\kappa =  \sqrt{(l + \frac{1}{2})^{2}  - (Z \alpha)^{2} - q^{2}},
\end{equation}
and $\sqrt{r e^{i\phi}} \equiv r^{1/2} e^{i\phi/2}$, $-\pi < \phi
\leq \pi$. Here and below we put the following restriction on $Z$:
$Z\alpha < \frac{1}{2}.$

The solution of (3.19) may be expressed in terms of hypergeometric functions
\begin{equation}
{v}(x) = d_{+} {F}(A_{+},B_{+},C_{+},x) +
 d_{-} x^{-2\kappa} {F}(A_{-},B_{-},C_{-},x),
\end{equation}
where $d_{+}, d_{-}$ are arbitrary constants and
\begin{eqnarray}
&&A_{\pm} = \pm \kappa  + \frac{1}{2} [\lambda + 1 -
\sqrt{\lambda^{2} + 4 Z \alpha (\varepsilon - Z \alpha)}],  \nonumber \\ 
&&B_{\pm} = \pm \kappa  + \frac{1}{2} [\lambda + 1 + 
\sqrt{\lambda^{2} + 4 Z \alpha (\varepsilon - Z \alpha)}], \nonumber  \\ 
&&C_{\pm} = \pm 2 \kappa  +  1 . \nonumber \end{eqnarray}

Using the asymptotic formulas for the hypergeometric functions \cite{NU} (for
$x \rightarrow 0$ and $x \rightarrow 1$), we find that the function 
$Q$, defined by (3.18) and (3.21), satisfies the restriction (3.15), 
if and only if $d_{-} = 0$  and
\begin{equation}
A_{+} = -n,
\end{equation}
$n = 0, 1, 2, \ldots $. In this case
\begin{equation}
{v}(x) = const {P_{n}^{(2\kappa, \lambda)}}(1- 2x),
\end{equation}
where ${P_{n}^{(\alpha, \beta)}}(z)$ is the Jacobi polynomial \cite{NU}
(see also Appendix).

Solving the equation (3.22), we get
\begin{equation}
\varepsilon = Z \alpha + N \frac{[\mu^{2} + 1- N^{2}-(Z \alpha)^{2}]^{1/2}}
{[N^{2}+(Z \alpha)^{2}]^{1/2}},
\end{equation}
where
\begin{equation}
N = n + \kappa + \frac{1}{2}
\end{equation}
is the principal quantum number satisfying the unequality
\begin{equation}
N < N_{0} \equiv 
(Z \alpha)^{1/2}[(\mu^{2} + 1)^{1/2}-Z \alpha]^{1/2}.
\end{equation}

Thus, there exists only a finite number of normalizable solutions of the 
equation of motion (2.15), that have the form (3.1) and are eigenfunctions
of the operators $(\vec{L}_{q})^{2}$ and $L_{q}^{3}$  . 
These solutions are the discrete spectrum solutions.

It follows from the definitions (3.20), (3.25) and the unequalitiy
(3.26) that the discrete spectrum is absent for $N_{0} \leq \frac{1}{2}$.
For $N_{0} > \frac{1}{2}$  it is also absent if
\begin{equation}
|q| \geq |q|_{0} = (N_{0})^{2} - N_{0} + (Z \alpha)^{2}
\end{equation}  
and exists, if $|q| < |q|_{0}$ . In this case $\varepsilon  =
{\varepsilon}(N) = {\varepsilon}({N}(n,l,|q|))$, 
where the principal quantum number $N$ is defined in (3.25) and
\begin{equation}
l = |q|, \ldots , {l_{0}}(|q|), \qquad n = 0, \ldots , {n_{0}}(l,|q|),
\end{equation}
In (3.28)
\begin{eqnarray}
{l_{0}}(|q|) \equiv max \{l| l- |q| = 0,1, \ldots ;
l(l +1) - q^{2} < |q|_{0}\}, \nonumber \\
{n_{0}}(l,|q|) \equiv max \{n|n = 0,1, \ldots ; n + \kappa + \frac{1}{2}
< N_{0} \} \nonumber
\end{eqnarray}
(the relations for $l_{0}$ and $n_{0}$ follow from the unequality (3.26)).

In the initial notations we have the following expression for the 
energy spectrum
\begin{equation}
E = \frac{Ze^{2}}{a} + N \frac{[m_{0}^{2} c^{4} + 
(1- N^{2}-(Z \alpha)^{2})(\hbar^{2}c^{2}/a^{2})]^{1/2}}
{[N^{2}+(Z \alpha)^{2}]^{1/2}},
\end{equation}
where $N < N_{0}(a)$, $ N_{0}(a) > 1/2$ and $|q| < |q|_{0}={|q|_{0}}(a)$.

Due to (3.29)
\begin{equation}
Ze^{2}/a < E < m_{0}c^{2}.
\end{equation}

The meson wave function, corresponding to the set of quantum numbers
$(n, l, m)$, is
\begin{equation}
\varphi = C \exp (-iEt/ \hbar )(\frac{2}{\coth \chi +1})^{
-\frac{1}{2} + \kappa} \exp[ -\chi (1+ \lambda)] 
 {P_{n}^{(2\kappa,\lambda)}}(\frac{\coth \chi -3}{\coth \chi +1})
Y_{qlm},
\end{equation}                    
where $C$ is constant and $n, l$ and $m$ satisfy the restrictions 
(3.28) and (3.6) correspondingly, $\kappa = {\kappa}(l, |q|)$ 
and $\lambda = {\lambda}(E,a)$ are defined in (3.20). 

Now we show that the parameter $E$ is  the
energy, corresponding to the meson wave function, appropriately normalized.
The energy functional, corresponding to the Lagrangian (2.16), is
\begin{eqnarray}
{\cal E}(\varphi, v) = &&\int_{U_{\ast}} d^{3}\vec{x}
(det \gamma_{ij})^{1/2}\{ \frac {\hbar^{2}}{c^{2}}\bar{v_{\ast}} v_{\ast} -
\frac {e^{2}}{c^{2}}V^{2} \overline{\varphi_{\ast}} \varphi_{\ast}
\nonumber  \\
&&+\hbar^{2} \gamma^{ij }\overline{({D_{i}}({\cal A}^{\ast})\varphi_{\ast})}
({D_{j}}({\cal A}^{\ast})\varphi_{\ast}) + 
m_{0}^{2}c^{2}\overline{\varphi_{\ast}} \varphi_{\ast}\}.
\end{eqnarray}
The energy is conserved on the solutions of the equation of motion (2.15):
${\cal E} = {\cal E}({\varphi}(t),{\dot{\varphi}}(t)) = const$.                                       
The Lagrangian (2.16) is invariant under the $U(1)$-transformations:
$\varphi \mapsto \varphi^{s} = \exp(-ise/\hbar) \varphi.$
Due to the E. Noether's theorem we have 
$Q = {Q}({\varphi}(t),{\dot{\varphi}}(t)) = const$, where                                       
\begin{equation}
{Q}(\varphi, v) = \int_{U_{\ast}} d^{3}\vec{x}(det \gamma_{ij})^{1/2}\{
i \hbar (\overline{\varphi_{\ast}} v_{\ast} - 
\overline{v_{\ast}} \varphi_{\ast})-2eV 
\overline{\varphi_{\ast}} \varphi_{\ast}\}
\end{equation}
is the charge functional ($Q : H \times H \longrightarrow R$).
Using (2.15) and (3.1), we get ${\cal E} = EQ/e$.
The physical normalization of the wave function 
$Q = {Q}({\varphi}(t),{\dot{\varphi}}(t)) = e$                                       
entails ${\cal E} = E$. So, $E$ is the energy of the scalar particle (meson).

Let us consider the flat-space limit:$a \rightarrow +\infty$ . 
In this case $ |q|_{0}, N_{0} \rightarrow +\infty$
and the discrete spectrum (3.29) contains an infinite number of levels 
for all values of $q$. For $q = 0$ and $a \rightarrow +\infty$
the formulas (3.29) and (3.31) coincide with the well-known relations 
(see, for example [14]).

For $a \sim 10^{28} cm$ (present cosmological scale), $Z = 1$ and           
$m = m_{\pi^{+}}$ (the mass of $\pi^{+}$-meson) we have: 
$N_{0} \sim 10^{20}$ and $|q|_{0} \sim 10^{40}$.

\begin{center}{\bf Appendix}   \end{center}

  Here we present the explicit expressions for the monopole 
spherical harmonics $Y_{qlm}$, $l = |q|, |q| + 1, \ldots$ ;
$ m = -l, -l+1, \ldots , l$ ; $Y_{qlm}$ are smooth sections of
the monopole vector $C$-bundle over the sphere $S^{2}$.
In the local trivialization over $S^{2}_{\pm} = S^{2} \setminus
\{\theta = \frac{\pi}{2} \pm \frac{\pi}{2}\}$ the sections $Y_{qlm}$
are represented by the complex-valued functions on $S^{2}_{\pm}$ 
\begin{equation}  
(Y_{qlm})_{\pm} = M_{qlm} {P_{n}^{(\alpha,\beta)}}(\cos \theta)
\exp(i(m \pm q) \varphi),   \nonumber   
\end{equation}
where  \begin{equation}
\alpha =-q-m, \qquad \beta = q -m, \qquad n=l+m \end{equation}
and ${P_{n}^{(\alpha,\beta)}}(x)$ is Jacobi polynomial
\begin{eqnarray}  
{P_{n}^{(\alpha ,\beta )}}(x) = &&\frac{(-1)^{n}}{2^{n} n!}
(1-x)^{-\alpha } (1+x)^{-\beta} \frac{d^{n}}{dx^{n}}
[(1-x)^{\alpha +n}(1+x)^{\beta +n}]    \nonumber \\
= &&\frac{{\Gamma}(n+\alpha +1)}{n!{\Gamma}(\alpha +1)}
{F}(-n, n+ \alpha + \beta +1,\alpha  +1, (1- x)/2), \nonumber
\end{eqnarray}
($M_{qlm}$ are constants).

\begin{center}{\bf Acknowledgments}   \end{center}

One of the authors (V. D. I.) is grateful to B. Allen and L. Parker for their 
hospitality at the University of Wisconsin-Milwaukee and useful discussions.
This work was supported in part by the Russian Ministry of Science
within the "Cosmomicrophysics" Project.

\pagebreak

{}

\end{document}